# RoxyBot-06: Stochastic Prediction and Optimization in TAC Travel

**Amy Greenwald**                                              AMY@CS.BROWN.EDU
*Department of Computer Science, Brown University*
*Providence, RI 02912 USA*

**Seong Jae Lee**                                    SEONGJAE@U.WASHINGTON.EDU
*Computer Science and Engineering, University of Washington*
*Seattle, WA 98195 USA*

**Victor Naroditskiy**                                        VNARODIT@CS.BROWN.EDU
*Department of Computer Science, Brown University*
*Providence, RI 02912 USA*

## Abstract

In this paper, we describe our autonomous bidding agent, RoxyBot, who emerged victorious in the travel division of the 2006 Trading Agent Competition in a photo finish. At a high level, the design of many successful trading agents can be summarized as follows: (i) price prediction: build a model of market prices; and (ii) optimization: solve for an approximately optimal set of bids, given this model. To predict, RoxyBot builds a stochastic model of market prices by simulating *simultaneous ascending auctions*. To optimize, RoxyBot relies on the *sample average approximation* method, a stochastic optimization technique.

## 1. Introduction

The annual Trading Agent Competition (TAC) challenges its entrants to design and build autonomous agents capable of effective trading in an online travel[1] shopping game. The first TAC, held in Boston in 2000, attracted 16 entrants from six countries in North America, Europe, and Asia. Excitement generated from this event led to refinement of the game rules, and continuation of regular tournaments with increasing levels of competition over the next six years. Year-by-year, entrants improved their designs, developing new ideas and building on previously successful techniques. Since TAC's inception, the lead author has entered successive modifications of her autonomous trading agent, RoxyBot. This paper reports on RoxyBot-06, the latest incarnation and the top scorer in the TAC-06 tournament.

The key feature captured by the TAC travel game is that goods are highly interdependent (e.g., flights and hotels must be coordinated), yet the markets for these goods operate independently. A second important feature of TAC is that agents trade via three different kinds of market mechanisms, each of which presents distinct challenges. Flights are traded in a posted-price environment, where a designated party sets a price that the other parties must "take or leave." Hotels are traded in simultaneous ascending auctions, like the FCC spectrum auctions. Entertainment tickets are traded in continuous double auctions, like the New York Stock Exchange. In grappling with all three mechanisms while constructing their agent strategies, participants are confronted by a number of interesting problems.

---

1. There are now four divisions of TAC: Travel, Supply Chain Management (SCM), CAT (TAC backwards), and Ad Auctions (AA). This paper is concerned only with the first; for a description of the others, see the papers by Arunachalam and Sadeh (2005), Cai et al. (2009), Jordan and Wellman (2009), respectively. In this paper, when we say TAC, we mean TAC Travel.

The success of an autonomous trading agent such as a TAC agent often hinges upon the solutions to two key problems: (i) *price prediction*, in which the agent builds a model of market prices; and (ii) *optimization*, in which the agent solves for an approximately optimal set of bids, given this model. For example, at the core of RoxyBot's 2000 architecture (Greenwald & Boyan, 2005) was a *deterministic* optimization problem, namely how to bid given price predictions in the form of point estimates. In spite of its effectiveness in the TAC-00 tournament, a weakness of the 2000 design was that RoxyBot could not explicitly reason about variance within prices. In the years since 2000, we recast the key challenges faced by TAC agents as several different *stochastic* bidding problems (see, for example, the paper by Greenwald & Boyan, 2004), whose solutions exploit price predictions in the form of distributions. In spite of our perseverance, RoxyBot fared unimpressively in tournament conditions year after year, until 2006. Half a decade in the laboratory spent searching for bidding heuristics that can exploit stochastic information at reasonable computational expense finally bore fruit, as RoxyBot emerged victorious in TAC-06. In a nutshell, the secret of RoxyBot-06's success is: (hotel) price prediction by simulating simultaneous ascending auctions, and optimization based on the sample average approximation method. Details of our approach are the subject of the present article.

**Overview** This paper is organized as follows. Starting in Section 2, we summarize the TAC market game. Next, in Section 3, we present a high-level view of RoxyBot's 2006 architecture. In Section 4, we describe RoxyBot's price prediction techniques for flights, hotels, and entertainment, in turn. Perhaps of greatest interest is our hotel price prediction method. Following Wellman et al. (2005), we predict hotel prices by computing approximate competitive equilibrium prices. Only, instead of computing those prices by running the tâtonnement process, we simulate simultaneous ascending auctions. Our procedure is simpler to implement than tâtonnement, yet achieves comparable performance, and runs sufficiently fast. In Section 5, we describe RoxyBot's optimization technique: sample average approximation. We argue that this approach is optimal in pseudo-auctions, an abstract model of auctions. In Section 6.1, we describe simulation experiments in a controlled testing environment which show that our combined approach—simultaneous ascending auctions for hotel price prediction and sample average approximation for bid optimization—performs well in practice in comparison with other reasonable bidding heuristics. In Section 6.2, we detail the results of the TAC-06 tournament, further validating the success of RoxyBot-06's strategy, and reporting statistics that shed light on the bidding strategies of other participating agents. Finally, in Section 7, we evaluate the collective behavior of the autonomous agents in the TAC finals since 2002. We find that the accuracy of competitive equilibrium calculations has varied from year to year and is highly dependent on the particular agent pool. Still, generally speaking, the collective appears to be moving toward competitive equilibrium behavior.

## 2. TAC Market Game: A Brief Summary

In this section, we summarize the TAC game. For more details, see `http://www.sics.se/tac/`.

Eight agents play the TAC game. Each is a simulated travel agent whose task is to organize itineraries for its clients to travel to and from "TACTown" during a five day (four night) period. In the time allotted (nine minutes), each agent's objective is to procure travel goods as inexpensively as possible, trading off against the fact that those goods are ultimately compiled into feasible trips that satisfy its client preferences to the greatest extent possible. The agents know the preferences of their own eight clients only, not the other 56.

Travel goods are sold in simultaneous auctions as follows:

- Flight tickets are sold by "TACAir" in dynamic posted-pricing environments. There are flights both to and from TACTown on each applicable day. No resale of flight tickets by agents is permitted.

  Flight price quotes are broadcast by the TAC server every ten seconds.

- Hotel reservations are sold by the "TAC seller" in multi-unit ascending call markets. Specifically, 16 hotel reservations are sold in each hotel auction to the 16 highest bidders at the $16th$ highest price. The TAC seller runs eight hotel auctions, one per night-hotel combination (recall that travel takes place during a four night period; moreover, there are two hotels: a good one and a bad one). No resale of hotel reservations by agents is permitted. Nor is bid withdrawal allowed.

  More specifically, the eight hotel auctions clear on the minute with exactly one auction closing at each of minutes one through eight. (The precise auction to close is chosen at random, with all open auctions equally likely to be selected.) For the auction that closes, the TAC server broadcasts the final closing price, and informs each agent of its winnings. For the others, the TAC server reports the current ask price, and informs each agent of its "hypothetical quantity won" (HQW).

- Agents are allocated an initial endowment of entertainment tickets, which they trade among themselves in continuous double auctions (CDAs). There are three entertainment events scheduled each day.

  Although the event auctions clear continuously, price quotes are broadcast only every 30 seconds.

One of the primary challenges posed by TAC is to design and build autonomous agents that bid effectively on interdependent (i.e., complementary or substitutable) goods that are sold in separate markets. Flight tickets and hotel reservations are complementary because flights are not useful to a client without the corresponding hotel reservations, nor vice versa. Tickets to entertainment events (e.g., the Boston Red Sox and the Boston Symphony Orchestra) are substitutable because a client cannot attend multiple events simultaneously.

```
REPEAT
      {start bid interval}
   0.  Download current prices and winnings from server
   1.  predict: build stochastic models
            a.  flights: Bayesian updating/learning
            b.  hotels: simultaneous ascending auctions
            c.  entertainment: sample historical data
   2.  optimize: sample average approximation
   3.  Upload current bids to server
       (three separate threads)
       {end bid interval}
UNTIL game over
```

Table 1: A high-level view of RoxyBot-06's architecture.

## 3. RoxyBot-06's Architecture: A High-Level View

In our approach to the problem of bidding on interdependent goods in the separate TAC markets, we adopt some simplifying assumptions. Rather than tackle the game-theoretic problem of characterizing strategic equilibria, we focus on a single agent's (decision-theoretic) problem of optimizing its own bidding behavior, assuming the other agents' strategies are fixed. In addition, we assume that the environment can be modeled in terms of the agent's predictions about market clearing prices. These prices serve to summarize the relevant information hidden in other agents' bidding strategies. These two assumptions—fixed other-agent behaviors and market information encapsulated by prices—support the modular design of RoxyBot-06 and many other successful TAC agents, which consists of two key stages: (i) price prediction; and (ii) optimization.

The optimization problem faced by TAC agents is a dynamic one that incorporates aspects of sequentiality as well as simultaneity in auctions. The markets operate simultaneously, but in addition, prices are discovered incrementally over time. In principle, a clairvoyant agent—one with knowledge of future clearing prices—could justifiably employ an open-loop strategy: it could solve the TAC optimization problem once at the start of the game and place all its bids accordingly, never reconsidering those decisions. A more practical alternative (and the usual approach taken in TAC[2]), is to incorporate into an agent's architecture a closed loop, or *bidding cycle*, enabling the agent to condition its behavior on the evolution of prices. As price information is revealed, the agent improves its price predictions, and reoptimizes its bidding decisions, repeatedly.

One distinguishing feature of RoxyBot-06 is that it builds stochastic models of market clearing prices, rather than predicting clearing prices as point estimates. Given its stochastic price predictions, stochastic optimization lies at the heart of RoxyBot-06. Assuming time is discretized into stages, or bid intervals, during each iteration of its bidding cycle, RoxyBot-06 faces an $n$-stage stochastic optimization problem, where $n$ is the number of stages remaining in the game. The key input to this optimization problem is a sequence of $n-1$ stochastic models of future prices, each one a joint probability distribution over all goods conditioned on past prices and past hotel closings. The solution to this optimization problem, and the output of each iteration of the bidding cycle, is a vector of bids, one per good (or auction).

---

2. An exception is livingagents (Fritschi & Dorer, 2002), the winner of TAC 2001.

Table 1 presents a high-level view of RoxyBot-06's architecture, emphasizing its bidding cycle. At the start of each bid interval, current prices and winnings are downloaded from the TAC server. Next, the key prediction and optimization routines are run. In the prediction module, stochastic models of flight, hotel, and entertainment prices are built. In the optimization module, bids are constructed as an approximate solution to an $n$-stage stochastic optimization problem. Prior to the end of each bid interval, the agents' bids are uploaded to the TAC server using three separate threads: (i) the flight thread bids on a flight only if its price is near its predicted minimum; (ii) the hotel thread bids on open hotels only if it is moments before the end of a minute; and (iii) the entertainment thread places bids immediately.

We discuss the details of RoxyBot-06's price prediction module first, and its optimization module second.

## 4. Price Prediction

In this section, we describe how RoxyBot-06 builds its stochastic models of flight, hotel, and event prices. Each model is a discrete probability distribution, represented by a set of "scenarios." Each scenario is a vector of "future" prices—prices at which goods can be bought and sold after the current stage. For flights, the price prediction model is not stochastic: the future buy price is simply RoxyBot-06's prediction of the expected minimum price during the current stage. For hotels, the future buy prices are predicted by Monte Carlo simulations of simultaneous ascending auctions to approximate competitive equilibrium prices. There are no current buy prices for hotels. For entertainment, RoxyBot-06 predicts future buy and sell prices based on historical data. Details of these price prediction methods are the focus of this section.

### 4.1 Flights

Efforts to deliberate about flight purchasing start with understanding the TAC model of flight price evolution.

#### 4.1.1 TAC FLIGHT PRICES' STOCHASTIC PROCESS

Flight prices follow a biased random walk. They are initialized uniformly in the range $[250, 400]$, and constrained to remain in the range $[150, 800]$. At the start of each TAC game instance, a bound $z$ on the final perturbation value is selected for each flight. These bounds are not revealed to the agents. What is revealed to the agents is a sequence of random flight prices. Every ten seconds, TACAir perturbs the price of each flight by a random value that depends on the hidden parameter $z$ and the current time $t$ as follows: given constants $c, d \in \mathbb{R}$ and $T > 0$, each (intermediate) bound on the perturbation value is a linear function of $t$:

$$x(t, z) = c + \frac{t}{T}(z - c) \qquad (1)$$

The perturbation value at time $t$ is drawn uniformly from one of the following ranges (see Algorithm 1):

- $U[-c, x(t, z)]$, if $x(t, z) > 0$

- $U[-c, +c]$, if $x(t, z) = 0$

- $U[x(t, z), +c]$, if $x(t, z) < 0$

Observe that the expected perturbation value in each case is simply the average of the corresponding upper and lower bounds. In particular,

- if $x(t, z) > c$, then the expected perturbation is positive;

- if $x(t, z) \in (0, c)$, then the expected perturbation is negative;

- if $x(t, z) \in (-c, 0)$, then the expected perturbation is positive;

- otherwise, if $x(t, z) \in \{-c, 0, c\}$, then the expected perturbation is zero.

Moreover, using Equation 1, we can compute the expected perturbation value conditioned on $z$:

- if $z \in [0, c]$, then $x(t, z) \in [0, c]$, so prices are expected not to increase;

- if $z \in [c, c + d]$, then $x(t, z) \in [c, c + d]$, so prices are expected not to decrease;

- if $z \in [-c, 0]$, then $x(t, z) \in [-c, c]$, so prices are expected not to increase while $t \leq \frac{cT}{c-z}$ and they are expected not to decrease while $t \geq \frac{cT}{c-z}$.

The TAC parameters are set as follows: $c = 10$, $d = 30$, $T = 540$, and $z$ uniformly distributed in the range $[-c, d]$. Based on the above discussion, we note the following: given no further information about $z$, TAC flight prices are expected to increase (i.e., the expected perturbation is positive); however, conditioned on $z$, TAC flight prices may increase *or* decrease (i.e., the expected perturbation can be positive or negative).

### 4.1.2 RoxyBot-06's Flight Prices Prediction Method

Although the value of the hidden parameter $z$ is never revealed to the agents, recall that the agents do observe sample flight prices, say $y_1, \ldots, y_t$, that depend on this value. This information can be used to model the probability distribution $P_t[z] \equiv P[z \mid y_1, \ldots, y_t]$. Such a probability distribution can be estimated using Bayesian updating. Before RoxyBot-06, agents Walverine (Cheng et al., 2005) and Mertacor (Toulis et al., 2006) took this approach. Walverine uses Bayesian updating to compute the next expected price perturbation and then compares that value to a threshold, postponing its flight purchases if prices are not expected to increase by more than that threshold. Mertacor uses Bayesian updating to estimate the time at which flight prices will reach their minimum value. RoxyBot uses Bayesian updating to compute the expected minimum price, as we now describe.

---

**Algorithm 1** getRange($c, t, z$)

---

compute $x(t, z)$ {Equation 1}
**if** $x(t, z) > 0$ **then**
$\quad a = -c$; $b = \lceil x(t, z) \rceil$
**else if** $x(t, z) < 0$ **then**
$\quad a = \lfloor x(t, z) \rfloor$; $b = +c$
**else**
$\quad a = -c$; $b = +c$
**end if**
**return** $[a, b]$ {range}

---

RoxyBot-06's implementation of Bayesian updating is presented in Algorithm 2. Letting $Q_0[z] = \frac{1}{c+d} = P[z]$, the algorithm estimates $P_{t+1}[z] = P[z \mid y_1, \ldots, y_{t+1}]$ as usual:

$$P[z \mid y_1, \ldots, y_t] = \frac{P[y_1, \ldots, y_t \mid z]P[z]}{\sum_{z'} P[y_1, \ldots, y_t \mid z']P[z']\,dz'} \tag{2}$$

where

$$P[y_1, \ldots, y_t \mid z] = \prod_{i=1}^{t} P[y_i \mid y_1, \ldots, y_{i-1}, z] \tag{3}$$

$$= \prod_{i=1}^{t} P[y_i \mid z] \tag{4}$$

Equation 4 follows from the fact that future observations are independent of past observations; observations depend only on the hidden parameter $z$.

The only thing left to explain is how to set the values $P[y_i \mid z]$, for $i = 1, \ldots, t$. As described in the pseudocode, this is done as follows: if $y_{t+1}$ is within the appropriate range at that time, then this probability is set uniformly within the bounds of that range; otherwise, it is set to 0. Presumably, Walverine's and Mertacor's implementations of Bayesian updating are not very different from this one.[3] However, as alluded to above, how the agents make use of the ensuing estimated probability distributions does differ.

RoxyBot-06 predicts each flight's price to be its expected minimum price. This value is computed as follows (see Algorithm 3): for each possible value of the hidden parameter $z$, RoxyBot simulates an "expected" random walk, selects the minimum price along this walk, and then outputs as its prediction the expectation of these minima, averaging according to $P_t[z]$. We call this random walk "expected," since the perturbation value $\Delta$ is an expectation (i.e., $\Delta = \frac{b-a}{2}$) instead of a sample (i.e., $\Delta \sim U[a, b]$). By carrying out this computation, RoxyBot generates flight price predictions that are point estimates. The implicit decision to make only RoxyBot-06's hotel and event price predictions stochastic was made based on our intuitive sense of the time vs. accuracy tradeoffs in RoxyBot's optimization module, and hence warrants further study.

---

3. We provide details here, because corresponding details for the other agents do not seem to be publicly available.

---

**Algorithm 2** Flight_Prediction$(c, d, t, y_{t+1}, Q_t)$

    **for all** $z \in \{-c, -c+1, \ldots, d\}$ **do**
       $[a, b] = \text{getRange}(c, t, z)$
       **if** $y_{t+1} \in [a, b]$ **then**
          $P[y_{t+1} \mid z] = \frac{1}{b-a}$
       **else**
          $P[y_{t+1} \mid z] = 0$
       **end if**
       $Q_{t+1}[z] = P[y_{t+1} \mid z] Q_t[z]$
    **end for**{update probabilities}
    **for all** $z \in \{-c, -c+1, \ldots, d\}$ **do**
       $P_{t+1}[z] = \frac{Q_{t+1}[z]}{\sum_{z'} Q_{t+1}[z'] \, dz'}$
    **end for**{normalize probabilities}
    **return** $P_{t+1}$ {probabilities}

---

**Algorithm 3** Expected_Minimum_Price$(c, t, t', p_t, P_t)$

    **for all** $z \in R$ **do**
       $\min[z] = +\infty$
       **for** $\tau = t+1, \ldots, t'$ **do**
          $[a, b] = \text{getRange}(c, \tau, z)$
          $\Delta = \frac{b-a}{2}$ {expected perturbation}
          $p_\tau = p_{\tau-1} + \Delta$ {perturb price}
          $p_\tau = \max(150, \min(800, p_\tau))$
          **if** $p_\tau < \min[z]$ **then**
             $\min[z] = p_\tau$
          **end if**
       **end for**
    **end for**
    **return** $\sum_z P_t[z] \min[z] \, dz$

---

## 4.2 Hotels

In a competitive market where each individual's effect on prices is negligible, equilibrium prices are prices at which supply equals demand, assuming all producers are profit-maximizing and all consumers are utility-maximizing. RoxyBot-06 predicts hotel prices by simulating *simultaneous ascending auctions* (SimAA) (Cramton, 2006), in an attempt to approximate competitive equilibrium (CE) prices. This approach is inspired by Walverine's (Cheng et al., 2005), where the tâtonnement method (Walras, 1874) is used for the same purpose.

### 4.2.1 Simultaneous Ascending Auctions

Let $\vec{p}$ denote a vector of prices. If $\vec{y}(\vec{p})$ denotes the cumulative supply of all producers, and if $\vec{x}(\vec{p})$ denotes the cumulative demand of all consumers, then $\vec{z}(\vec{p}) = \vec{x}(\vec{p}) - \vec{y}(\vec{p})$ denotes the excess demand in the market. The tâtonnement process adjusts the price vector at iteration $n + 1$, given the price vector at iteration $n$ and an adjustment rate $\alpha_n$ as follows: $\vec{p}_{n+1} = \vec{p}_n + \alpha_n \vec{z}(\vec{p}_n)$. SimAA adjusts the price vector as follows: $\vec{p}_{n+1} = \vec{p}_n + \alpha \max\{\vec{z}(\vec{p}_n), 0\}$, for some fixed value of $\alpha$. Both of these processes continue until excess demand is non-positive: i.e., supply exceeds demand.

Although competitive equilibrium prices are not guaranteed to exist in TAC markets (Cheng et al., 2003), the SimAA adjustment process, is still guaranteed to converge: as prices increase, demand decreases while supply increases; hence, supply eventually exceeds demand. The only parameter to the SimAA method is the magnitude $\alpha$ of the price adjustment. The smaller this value, the more accurate the approximation (assuming CE prices exist), so the value of $\alpha$ can be chosen to be the lowest value that time permits.

The tâtonnement process, on the other hand, is more difficult to apply as it is not guaranteed to converge. The Walverine team dealt with the convergence issue by decaying an initial value of $\alpha$. However, careful optimization was required to ensure convergence to a reasonable solution in a reasonable amount of time. In fact, Walverine found it helpful to set initial prices to certain non-zero values. This complexity is not present when using simultaneous ascending auctions to approximate competitive equilibrium prices.

### 4.2.2 Prediction Quality

In TAC, cumulative supply is fixed. Hence, the key to computing excess demand is to compute cumulative demand. Each TAC agent knows the preferences of its own clients, but must estimate the demand of the others. Walverine computes a single hotel price prediction (a point estimate) by considering its own clients' demands together with those of 56 "expected" clients. Briefly, the utility of an expected client is an average across travel dates and hotel types augmented with fixed entertainment bonuses that favor longer trips (see the paper by Cheng et al., 2005, for details). In contrast, RoxyBot-06 builds a stochastic model of hotel prices consisting of $S$ scenarios by considering its own clients' demands together with $S$ random samples of 56 clients. A (random or expected) client's demand is simply the quantity of each good in its optimal package, given current prices. The cumulative demand is the sum total of all client's individual demands.

In Figure 1, we present two scatter plots that depict the quality of various hotel price predictions at the beginning of the TAC 2002 final games. All price predictions are evaluated using two metrics: Euclidean distance and the "expected value of perfect prediction" (EVPP). Euclidean distance is a measure of the difference between two vectors, in this case the actual and the predicted prices. The value of perfect prediction (VPP) for a client is the difference between its surplus (value of its preferred package less price) based on actual and predicted prices. EVPP is the VPP averaged over the distribution of client preferences.[4]

On the left, we plot the predictions generated using the competitive equilibrium approximation methods, tâtonnement and SimAA, *both* with fixed $\alpha = \frac{1}{24}$, making expected, random, and exact predictions. The "exact" predictions are computed based on the actual clients in the games, not just the client distribution; hence, they serve as a lower bound on the performance of these techniques on this data set. Under both metrics, and for both expected and random, SimAA's predictions outperform tâtonnement's.

---

4. See the paper by Wellman et al. (2004) for details.

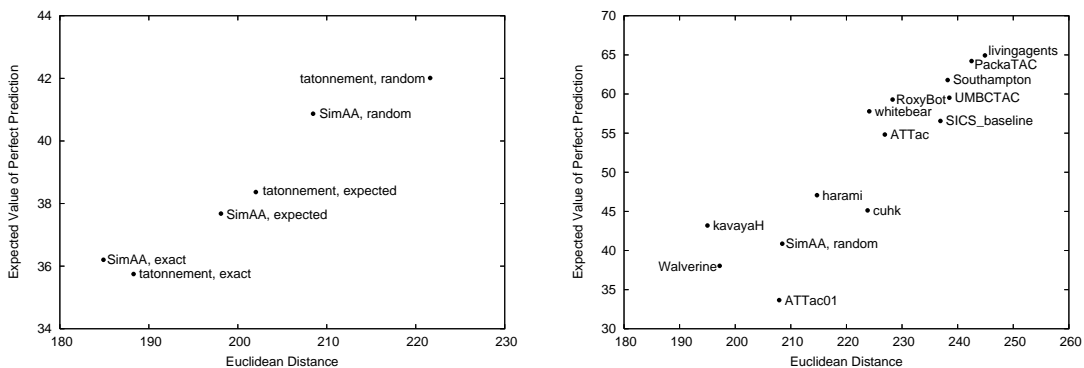

Figure 1: EVPP and Euclidean Distance for the CE price prediction methods (tâtonnement and SimAA with $\alpha = \frac{1}{24}$; expected, random, and exact) and the TAC 2002 agents' predictions in the 2002 finals (60 games). The plot on the left shows that SimAA's predictions are better than tâtonnement's and that expected's are better than random's. RoxyBot-06's method of hotel price prediction (SimAA, Random) is plotted again on the right. Note the differences in scales between the two plots.

Since $\alpha$ is fixed, and tâtonnement is not guaranteed to converge under this condition, this outcome is not entirely surprising. What is interesting, though, is that SimAA expected performs comparably to Walverine (see the right plot).[5] This is interesting because SimAA has fewer parameter settings than tâtonnement—only a single $\alpha$ value as compared to an initial $\alpha$ value together with a decay schedule—and moreover, we did not optimize its parameter setting. Walverine's parameter settings, on the other hand, were highly optimized.

We interpret each prediction generated using randomly sampled clients as a sample scenario, so that a set of such scenarios represents draws from a probability distribution over CE prices. The corresponding vector of predicted prices that is evaluated is actually the average of multiple (40) such predictions; that is, we evaluate an estimate of the mean of this probability distribution. The predictions generated using sets of random clients are not as good as the predictions with expected clients (see Figure 1 left), although with more than 40 sets of random clients, the results might improve. Still, the predictions with random clients comprise RoxyBot-06's stochastic model of hotel prices, which is key to its bidding strategy. Moreover, using random clients helps RoxyBot-06 make better interim predictions later in the game as we explain next.

### 4.2.3 PREDICTION QUALITY OVER TIME: INTERIM PRICE PREDICTION

The graphs depicted in Figure 1 pertain to hotel price predictions made at the beginning of the game, when all hotel auctions are open. In those CE computations, prices are initialized to 0. As hotel auctions close, RoxyBot-06 updates the predicted prices of the hotel auctions that remain open. We experimented with two ways of constructing interim price predictions. The first is to initialize and lower bound the prices in the hotel markets at their closing (for closed auctions) or current ask (for open auctions) prices while computing competitive equilibrium prices.[6] The second differs in its treatment of closed auctions: we simulate a process of distributing the goods in the closed auctions to the clients who want them most, and then exclude the closed markets (i.e., fix prices at $\infty$) from further computations of competitive equilibrium prices.

---

5. With the exception of the RoxyBot-06 data point (i.e., SimAA random), this plot was produced by the Walverine team (Wellman et al., 2004).

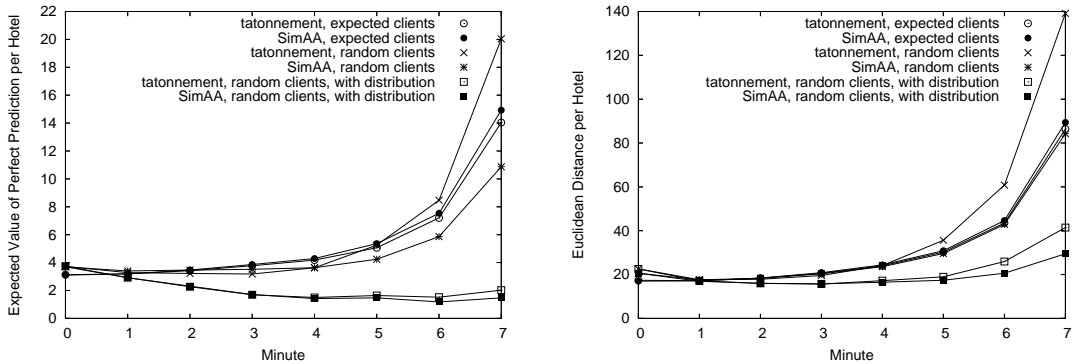

Figure 2: EVPP and Euclidean Distance in TAC 2006 finals (165 games) of the CE price prediction methods with and without distribution as the game progresses. Distribution improves prediction quality.

Regarding the second method—the distribution method—we determine how to distribute goods by computing competitive equilibrium prices again. As explained in Algorithm 4, all hotels (in both open and closed auctions) are distributed to *random* clients by determining who is willing to pay the competitive equilibrium prices for what. It is not immediately obvious how to distribute goods to expected clients; hence, we enhanced only the prediction methods with random clients with distribution.

Figure 2, which depicts prediction quality over time, shows that the prediction methods enhanced with distribution are better than the predictions obtained by merely initializing the prices of closed hotel auctions at their closing prices. Hotels that close early tend to sell for less than hotels that close late; hence, the prediction quality of any method that makes decent initial predictions is bound to deteriorate if those predictions remain relatively constant throughout the game.

### 4.2.4 Run Time

Table 2 shows the run times of the CE prediction methods on the TAC 2002 (60 games) and TAC 2006 (165 games) finals data set at minute 0, as well as their run times during

---

6. At first blush, it may seem more sensible to *fix* the prices of closed hotels at their closing prices, rather than merely lower bound them (i.e., allow them to increase). If some hotel closed at an artificially low price, however, and if that price were not permitted to increase, then the predicted prices of the hotels complementing the hotel in question would be artificially high.

---
**Algorithm 4** Distribute
---
1: **for all** hotel auctions $h$ **do**
2:     initialize price to 0
3:     initialize supply to 16
4: **end for**
5: compute competitive equilibrium prices {Tâtonnement or SimAA}
6: **for all** closed hotel auctions $h$ **do**
7:     distribute units of $h$ to those who demand them at the computed competitive equilibrium prices
8:     distribute any leftover units of $h$ uniformly at random
9: **end for**
---

minutes 1–7 on the TAC 2006 finals data set. What the numbers in this table convey is that SimAA's run time, even with distribution, is reasonable. For example, at minute 0, SimAA sample takes on the order of 0.1 seconds. At minutes 1-7, this method without distribution runs even faster. This speed increase occurs because CE prices are bounded below by current ask prices and above by the maximum price a client is willing to pay for a hotel, and current ask prices increase over time, correspondingly reducing the size of the search space. SimAA sample with distribution at minutes 1-7 takes twice as long as SimAA sample without distribution at minute 0 because of the time it takes to distribute goods, but the run time is still only (roughly) 0.2 seconds. Our implementation of tâtonnement runs so slowly because we fixed $\alpha$ instead of optimizing the tradeoff between convergence rate and accuracy, so the process did not converge, and instead ran for the maximum number of iterations (10,000). In summary, SimAA is simpler than tâtonnement to implement, yet performs comparably to an optimized version of tâtonnement (i.e., Walverine), and runs sufficiently fast.

|  | Exp Tât | Exp SimAA | Sam Tât | Sam SimAA | Dist Tât | Dist SimAA |
|---|---|---|---|---|---|---|
| 2002, minute 0 | 2213 | 507 | 1345 | 157 | — | — |
| 2006, minute 0 | 2252 | 508 | 1105 | 130 | 1111 | 128 |
| 2006, average 1–7 | 2248 | 347 | 1138 | 97 | 2249 | 212 |

Table 2: Run times for the CE price prediction methods, in milliseconds. Experiments were run on AMD Athlon(tm) 64 bit 3800+ dual core processors with 2M of RAM.

### 4.2.5 Summary

The simulation methods discussed in this section—the tâtonnement process and simultaneous ascending auctions—were employed to predict hotel prices only. (In our simulations, flight prices are fixed at their expected minima, and entertainment prices are fixed at 80.) In principle, competitive equilibrium (CE) prices could serve as predictions in all TAC markets. However, CE prices are unlikely to be good predictors of flight prices, since flight prices are determined exogenously. With regard to entertainment tickets, CE prices might have predictive power; however, incorporating entertainment tickets into the tâtonnement and SimAA calculations would have been expensive. (In our simulations, following Wellman et al., 2004, client utilities are simply augmented with fixed entertainment bonuses that favor longer trips.) Nonetheless, in future work, it could be of interest to evaluate the success of these or related methods in predicting CDA clearing prices.

Finally, we note that we refer to our methods of computing excess demand as "client-based" because we compute the demands of each client on an individual basis. In contrast, one could employ an "agent-based" method, whereby the demands of agents, not clients, would be calculated. Determining an agent's demands involves solving so-called *completion*, a deterministic (prices are known) optimization problem at the heart of RoxyBot-00's architecture (Greenwald & Boyan, 2005). As TAC completion is NP-hard, the agent-based method of predicting hotel prices is too expensive to be included in RoxyBot-06's inner loop. In designing RoxyBot-06, we reasoned that an architecture based on a stochastic pricing model generated using the client-based method and randomly sampled clients would outperform one based on a point estimate pricing model generated using the agent-based method and some form of expected clients, but we did not verify our reasoning empirically.

### 4.3 Entertainment

During each bid interval, RoxyBot-06 predicts current and future buy and sell prices for tickets to all entertainment events. These price predictions are optimistic: the agent assumes it can buy (or sell) goods at the least (or most) expensive prices that it expects to see before the end of the game. More specifically, each current price prediction is the best predicted price during the current bid interval.

RoxyBot-06's estimates of entertainment ticket prices are based on historical data from the past 40 games. To generate a scenario, a sample game is drawn at random from this collection, and the sequences of entertainment bid, ask, and transaction prices are extracted. Given such a history, for each auction $a$, let $trade_{ai}$ denote the price at which the last trade before time $i$ transacted; this value is initialized to 200 for buying and 0 for selling. In addition, let $bid_{ai}$ denote the bid price at time $i$, and let $ask_{ai}$ denote the ask price at time $i$.

RoxyBot-06 predicts the future buy price in auction $a$ after time $t$ as follows:

$$future\_buy_{at} = \min_{i=t+1,\ldots,T} \min\{trade_{ai}, ask_{ai}\} \tag{5}$$

In words, the future buy price at each time $i = t+1, \ldots, T$ is the minimum of the ask price after time $i$ and the most recent trade price. The future buy price at time $t$ is the minimum across the future buy prices at all later times. The future sell price after time $t$ is predicted analogously:

$$future\_sell_{at} = \max_{i=t+1,\ldots,T} \max\{trade_{ai}, bid_{ai}\} \tag{6}$$

Arguably, RoxyBot-06's entertainment predictions are made in the simplest possible way: past data are future predictions. It is likely one could improve upon this naive approach by using a generalization technique capable of "learning" a distribution over these data, and then sampling from the learned distribution.

### 4.4 Summary

In this section, we described RoxyBot-06's price prediction methods. The key ideas, which may be transferable if not beyond TAC, at least to other TAC agents, are as follows:

1. RoxyBot makes stochastic price predictions. It does so by generating a set of so-called "scenarios," where each scenario is a vector of future prices.

2. For each flight, RoxyBot uses Bayesian updating to predict its expected minimum price.

3. For hotels, RoxyBot-uses a method inspired by Walverine's: it approximates competitive equilibrium prices by simulating simultaneous ascending auctions, rather than the usual tâtonnement process.

## 5. Optimization

Next, we characterize RoxyBot-06's optimization routine. It is (i) stochastic, (ii) global, and (iii) dynamic. It takes as input stochastic price predictions; it considers its flight, hotel, and entertainment bidding decisions in unison; and it simultaneously reasons about bids to be placed in both current and future stages of the game.

### 5.1 Abstract Auction Model

Recall that our treatment of bidding is decision-theoretic, rather than game-theoretic. In particular, we focus on a single agent's problem of optimizing its own bidding behavior, assuming the other agents' strategies are fixed. In keeping with our basic agent architecture, we further assume that the environment can be modeled in terms of the agent's predictions about market clearing prices. We introduce the term *pseudo-auction* to refer to a market mechanism defined by these two assumptions—fixed other-agent behaviors and market information encapsulated by prices. The optimization problem that RoxyBot solves is one of bidding in pseudo-auctions, not (true) auctions. In this section, we formally develop this abstract auction model and relate it to TAC auctions; in the next, we define and propose heuristics to solve various pseudo-auction bidding problems.

#### 5.1.1 Basic Formalism

In this section, we formalize the basic concepts needed to precisely formulate bidding under uncertainty as an optimization problem, including: packages—sets of goods, possibly multiple units of each; a function that describes how much the agent values each package; pricelines—data structures in which to store the prices of each unit of each good; and bids—pairs of vectors corresponding to buy and sell offers.

**Packages** Let $G$ denote an ordered set of $n$ distinct goods and let $N \in \mathbb{N}^n$ represent the multiset of these goods in the marketplace, with $N_g$ denoting the number of units of each good $g \in G$. A *package* $M$ is a collection of goods, that is, a "submultiset" of $N$. We write $M \subseteq N$ whenever $M_g \leq N_g$ for all $g \in G$.

It is instructive to interpret this notation in the TAC domain. The flights, hotel rooms, and entertainment events up for auction in TAC comprise an ordered set of 28 distinct goods. In principle, the multiset of goods in the TAC marketplace is:

$$N^{\text{TAC}} = \langle \underbrace{\infty, \dots, \infty}_{8 \text{ flights}}, \underbrace{16, \dots, 16}_{8 \text{ hotels}}, \underbrace{8, \dots, 8}_{12 \text{ events}} \rangle \in \mathbb{N}^{28}$$

In practice, however, since each agent works to satisfy the preferences of only eight clients, it suffices to consider the multiset of goods:

$$N^{\text{TAC8}} = \langle \underbrace{8 \dots, 8}_{8 \text{ flights}}, \underbrace{8, \dots, 8}_{8 \text{ hotels}}, \underbrace{8, \dots, 8}_{12 \text{ events}} \rangle \subseteq N^{\text{TAC}}$$

A trip corresponds to a package, specifically some $M \subseteq N^{\text{TAC8}}$ that satisfies the TAC feasibility constraints.

Given $A, B \subseteq N$, we rely on the two basic operations, $\oplus$ and $\ominus$, defined as follows: for all $g \in G$,

$$(A \oplus B)_g \equiv A_g + B_g$$
$$(A \ominus B)_g \equiv A_g - B_g$$

For example, if $G = \{\alpha, \beta, \gamma\}$ and $N = \langle 1, 2, 3 \rangle$, then $A = \langle 0, 1, 2 \rangle \subseteq N$ and $B = \langle 1, 1, 1 \rangle \subseteq N$. Moreover, $(A \oplus B)_\alpha = 1$, $(A \oplus B)_\beta = 2$, and $(A \oplus B)_\gamma = 3$; and $(A \ominus B)_\alpha = -1$, $(A \ominus B)_\beta = 0$, and $(A \ominus B)_\gamma = 1$.

**Value** Let $\mathcal{N}$ denote the set of all submultisets of $N$: i.e., packages comprised of the goods in $N$. We denote $v : \mathcal{N} \rightarrow \mathbb{R}$ a function that describes the value the bidding agent attributes to each viable package.

In TAC, each agent's objective is to compile packages for $m = 8$ individual clients. As such, the agent's value function takes special form. Each client $c$ is characterized by its own value function $v_c : \mathcal{N} \rightarrow \mathbb{R}$, and the agent's value for a collection of packages is the sum of its clients' respective values for those packages: given a vector of packages $\vec{X} = (X_1, \dots, X_m)$,

$$v(\vec{X}) = \sum_{c=1}^{m} v_c(X_c). \tag{7}$$

**Pricelines** A *buyer priceline* for good $g$ is a vector $\vec{p}_g \in \mathbb{R}_+^{N_g}$, where the $k$th component, $p_{gk}$, stores the *marginal cost* to the agent of acquiring the $k$th unit of good $g$. For example, if an agent currently holds four units of a good $\tilde{g}$, and if four additional units of $\tilde{g}$ are available at costs of \$25, \$40, \$65, and \$100, then the corresponding buyer priceline (a vector of length 8) is given by $\vec{p}_{\tilde{g}} = \langle 0, 0, 0, 0, 25, 40, 65, 100 \rangle$. The leading zeros indicate that the four goods the agent holds may be "acquired" at no cost.

We assume buyer pricelines are nondecreasing. Note that this assumption is WLOG, since an optimizing agent buys cheaper goods before more expensive ones.

Given a set of buyer pricelines $P = \{\vec{p}_g \mid g \in G\}$, we define costs additively, that is, the *cost* of the goods in multiset $Y \subseteq N$ is given by:

$$\forall g, \quad \text{Cost}_g(Y, P) = \sum_{k=1}^{Y_g} p_{gk},$$
$$\text{Cost}(Y, P) = \sum_{g \in G} \text{Cost}_g(Y, P). \tag{8}$$

A *seller priceline* for good $g$ is a vector $\vec{\pi}_g \in \mathbb{R}_+^{N_g}$. Much like a buyer priceline, the $k$th component of a seller priceline for $g$ stores the *marginal revenue* that an agent could earn from the $k$th unit it sells. For example, if the market demands four units of good $\tilde{g}$, which can be sold at prices of \$20, \$15, \$10, and \$5, then the corresponding seller priceline is given by $\vec{\pi}_{\tilde{g}} = \langle 20, 15, 10, 5, 0, 0, 0, 0 \rangle$. Analogously to buyer pricelines, the tail of zero revenues indicates that the market demands only four of those units.

We assume seller pricelines are nonincreasing. Note that this assumption is WLOG, since an optimizing agent sells more expensive goods before cheaper ones.

Given a set of seller pricelines $\Pi = \{\vec{\pi}_g \mid g \in G\}$, we define revenue additively, that is, the *revenue* associated with multiset $Z \subseteq N$ is given by:

$$\forall g, \quad \text{Revenue}_g(Z, \Pi) \;=\; \sum_{k=1}^{Z_g} \pi_{gk}, \tag{9}$$

$$\text{Revenue}(Z, \Pi) \;=\; \sum_{g \in G} \text{Revenue}_g(Z, \Pi). \tag{10}$$

If a priceline is constant, we say that prices are *linear*. We refer to the constant value as a *unit price*. With linear prices, the cost of acquiring $k$ units of good $g$ is $k$ times the unit price of good $g$.

**Bids**  An agent submits a bid $\beta$ expressing offers to buy or sell various units of the goods in the marketplace. We divide $\beta$ into two components $\langle \vec{b}, \vec{a} \rangle$, where for each good $g$ the bid consists of a *buy offer*, $\vec{b}_g = \langle b_{g1}, \dots, b_{gN_g} \rangle$, and a *sell offer*, $\vec{a}_g = \langle a_{g1}, \dots, a_{gN_g} \rangle$. The bid price $b_{gk} \in \mathbb{R}_+$ (resp. $a_{gk} \in \mathbb{R}_+$) represents an offer to buy (sell) the $k$th unit of good $g$ at that price.

By definition, the agent cannot buy (sell) the $k$th unit unless it also buys (sells) units $1, \dots, k-1$. To accommodate this fact, we impose the following constraint: Buy offers must be nonincreasing in $k$, and sell offers nondecreasing. In addition, an agent may not offer to sell a good for less than the price at which it is willing to buy that good: i.e., $b_{g1} < a_{g1}$. Otherwise, it would simultaneously buy and sell good $g$. We refer to these restrictions as *bid monotonicity* constraints.

### 5.1.2 Pseudo-Auction Rules

Equipped with this formalism, we can specify the rules that govern pseudo-auctions. As in a true auction, the outcome of a pseudo-auction dictates the quantity of each good to exchange, and at what prices, conditional on the agent's bid. The quantity issue is resolved by the *winner determination rule* whereas the price issue is resolved by the *payment rule*.

**Definition 5.1** [Pseudo-Auction Winner Determination Rule] Given buyer and seller pricelines $P$ and $\Pi$, and bid $\beta = \langle \vec{b}, \vec{a} \rangle$, the agent buys the multiset of goods $\text{Buy}(\beta, P)$ and sells the multiset of goods $\text{Sell}(\beta, \Pi)$, where

$$\text{Buy}_g(\beta, P) = \max_k k \text{ such that } b_{gk} \geq p_{gk}$$

$$\text{Sell}_g(\beta, \Pi) = \max_k k \text{ such that } a_{gk} \leq \pi_{gk}$$

Note that the monotonicity restrictions on bids ensure that the agent's offer is better than or equal to the price for every unit it exchanges, and that the agent does not simultaneously buy and sell any good.

There are at least two alternative payment rules an agent may face. In a *first-price pseudo-auction*, the agent pays its bid price (for buy offers, or is paid its bid price for sell offers) for each good it wins. In a *second-price pseudo-auction*, the agent pays (or is paid) the prevailing prices, as specified by the realized buyer and seller pricelines. This terminology derives by analogy from the standard first- and second-price sealed bid auctions (Krishna, 2002; Vickrey, 1961). In these mechanisms, the high bidder for a single item pays its bid (the first price), or the highest losing bid (the second price), respectively. The salient property is that in first-price pseudo-auctions, the price is set by the bid of the winner, whereas in second-price pseudo-auctions an agent's bid price determines whether or not it wins but not the price it pays.

In this paper, we focus on the second-price model. That is, our basic problem definitions presume second-price auctions; however, our bidding heuristics are not tailored to this case. As in true auctions, adopting the second-price model in pseudo-auctions simplifies the problem for the bidder. It also provides a reasonable approximation to the situation faced by TAC agents, as we now argue:

- In TAC entertainment auctions, agents submit bids (i.e., buy and sell offers) of the form specified above. If we interpret an agent's buyer and seller pricelines as the current order book (not including the agent's own bid), then the agent's immediate winnings are as determined by the winner determination rule, and payments are according to the second-price rule (i.e., the order-book prices prevail).

- In TAC hotel auctions, only buy bids are allowed. Assuming once again an order book that reflects all outstanding bids other than the agent's own, an accurate buyer priceline would indicate that the agent can win $k$ units of a good if it pays—for *all* $k$ units—a price just above the $(17 - k)$th existing (other-agent) offer. The actual price it pays will be that of the 16th-highest unit offer (including its own offer). Since the agent's own bid may affect the price,[7] this situation lies between the first- and second-price characterizations of pseudo-auctions described above.

- In TAC flight auctions, agents may buy any number of units at the posted price. The situation at any given time is modeled exactly by the second-price pseudo-auction abstraction.

## 5.2 Bidding Problems

We are now ready to discuss the optimization module repeatedly employed by RoxyBot-06 within its bidding cycle to construct its bids. The key bidding decisions are: what goods to bid on, at what price, and when?

---

7. It can do so in two ways. First, the agent may submit the 16th-highest unit offer, in which case it sets the price. Second, when it bids for multiple units, the number it wins determines the price-setting unit, thus affecting the price for all winning units. Note that this second effect would be present even if the auction cleared at the 17th-highest price.

Although RoxyBot technically faces an $n$-stage stochastic optimization problem, it solves this problem by collapsing those $n$ stages into only two relevant stages, "current" and "future," necessitating only one stochastic model of future prices (current prices are known). This simplification is achieved by ignoring the potentially useful information that hotel auctions close one by one in a random, unspecified order, and instead operating (like most TAC agents) under the assumption that all hotel auctions close at the end of the current stage. Hence, there is only one model of hotel prices: a stochastic model of future prices. Moreover, the only pressing decisions regarding hotels are what goods to bid on now and at what price. There is no need to reason about the timing of hotel bid placement.

In contrast, since flight and entertainment auctions clear continuously, a trading agent should reason about the relevant tradeoffs in timing its placement of bids on these goods. Still, under the assumption that all hotel auctions close at the end of the current stage, in future stages, hotel prices, and hence hotel winnings, are known, so the only remaining decisions are what flight and entertainment tickets to buy. A rational agent will time its bids in these markets to capitalize on the "best" prices. (The best prices are the minima for buying and the maxima for selling.) Hence, it suffices for an agent's model of future prices in these markets to predict only the best prices (conditioned on current prices). That is, it suffices to consider only one stochastic pricing model. No further information is necessary.

Having established that it suffices for RoxyBot to pose and solve a two-stage, rather than an $n$-stage, stochastic optimization problem, we now proceed to define an abstract series of such problems that is designed to capture the essence of bidding under uncertainty in TAC-like hybrid markets that incorporate aspects of simultaneous and sequential, one-shot and continuously-clearing, auctions. More specifically, we formulate these problems as two-stage stochastic programs with integer recourse (see the book by Birge & Louveaux, 1997, for an introduction to stochastic programming).

In a two-stage stochastic program, there are two decision-making stages, and hence two sets of variables: first- and second-stage variables. The objective is to maximize the sum of the first-stage objectives (which depend only on the first-stage variables) and the expected value of the ensuing second-stage objectives (which can depend on both the first- and second-stage variables). The objective value in the second stage is called the *recourse* value, and if any of the second-stage variables are integer-valued, then the stochastic program is said to have *integer* recourse.

At a high-level, the bidding problem can be formulated as a two-stage stochastic program as follows: in the first stage, when current prices are known but future prices are uncertain, bids are selected; in the second stage, all uncertainty is resolved, and goods are exchanged. The objective is to maximize the expected value of the second-stage objective, namely the sum of the inherent value of final holdings and any profits earned, less any first-stage costs. Since the second stage involves integer-valued decisions (the bidder decides what goods to buy and sell at known prices), the bidding problem is one with integer recourse.

In this section, we formulate a series of bidding problems as two-stage stochastic programs with integer recourse, each one tailored to a different type of auction mechanism, illustrating a different type of bidding decision. The mechanisms we study, inspired by TAC, are one-shot and continuously-clearing variants of second-price pseudo-auctions. In the former, bids can only be placed in the first stage; in the latter, there is an opportunity for recourse. Ultimately, we combine all decision problems into one unified problem that captures what we mean by bidding under uncertainty.

In our formal problem statements, we rely on the following notation:

- Variables:

    - $Q^1$ is a multiset of goods to buy now
    - $Q^2$ is a multiset of goods to buy later
    - $R^1$ is a multiset of goods to sell now
    - $R^2$ is a multiset of goods to sell later

- Constants:

    - $P^1$ is a set of current buyer pricelines
    - $P^2$ is a set of future buyer pricelines
    - $\Pi^1$ is a set of current seller pricelines
    - $\Pi^2$ is a set of future seller pricelines

Note that $P^1$ and $\Pi^1$ are always known, whereas $P^2$ and $\Pi^2$ are uncertain in the first stage but their uncertainty is resolved in the second stage.

**Flight Bidding Problem**    An agent's task in bidding in flight auctions is to decide how many flights to buy now at current prices and later at the lowest future prices, given (known) current prices and a stochastic model of future prices. Although in TAC all units of each flight sell for the same price at any one time, we state the flight bidding problem more generally: we allow for different prices for different units of the same flight.

**Definition 5.2** [Continuously-Clearing, Buying] Given a set of current buyer pricelines $P^1$ and a probability distribution $f$ over future buyer pricelines $P^2$,

$$\mathsf{FLT}(f) = \max_{Q^1 \in \mathbb{Z}^n} \mathbb{E}_{P^2 \sim f} \left[ \max_{Q^2 \in \mathbb{Z}^n} v(Q^1 \oplus Q^2) - \left( \mathrm{Cost}(Q^1, P^1) + \mathrm{Cost}(Q^1 \oplus Q^2, P^2) - \mathrm{Cost}(Q^1, P^2) \right) \right] \tag{11}$$

Note that there are two cost terms referring to future pricelines ($\mathrm{Cost}(\cdot, P^2)$). The first of these terms adds the total cost of the goods bought in the first and second stages. The second term subtracts the cost of the goods bought in just the first stage. This construction ensures that, if an agent buys $k$ units of a good now, any later purchases of that good incur the charges of units $(k+1, k+2, ...)$ in the good's future priceline.

**Entertainment Bidding Problem**    Abstractly, the entertainment *buying* problem is the same as the flight bidding problem. An agent must decide how many entertainment tickets to buy now at current prices and later at the lowest future prices. The entertainment *selling* problem is the opposite of this buying problem. An agent must decide how many tickets to sell now at current prices and later at the highest future prices.

**Definition 5.3** [Continuously-Clearing, Buying and Selling] Given a set of current buyer and seller pricelines $(P, \Pi)^1$ and a probability distribution $f$ over future buyer and seller pricelines $(P, \Pi)^2$,

$$\mathsf{ENT}(f) = \max_{Q^1, R^1 \in \mathbb{Z}^n} \mathbb{E}_{(P,\Pi)^2 \sim f} \Big[ \max_{Q^2, R^2 \in \mathbb{Z}^n} v((Q^1 \oplus Q^2) \ominus (R^1 \oplus R^2))$$
$$- \big( \mathrm{Cost}(Q^1, P^1) + \mathrm{Cost}(Q^1 \oplus Q^2, P^2) - \mathrm{Cost}(Q^1, P^2) \big)$$
$$+ \big( \mathrm{Revenue}(R^1, \Pi^1) + \mathrm{Revenue}(R^1 \oplus R^2, \Pi^2) - \mathrm{Revenue}(R^1, \Pi^2) \big) \Big] \quad (12)$$

subject to $Q^1 \supseteq R^1$ and $Q^1 \oplus Q^2 \supseteq R^1 \oplus R^2$, for all $(P, \Pi)^2$.

The constraints ensure that an agent does not sell more units of any good than it buys.

**Hotel Bidding Problem**  Hotel auctions close at fixed times, but in an unknown order. Hence, during each iteration of an agent's bidding cycle, one-shot auctions approximate these auctions well. Unlike in the continuous setup, where decisions are made in both the first and second stages, in the one-shot setup, bids can only be placed in the first stage; in the second stage, winnings are determined and evaluated.

**Definition 5.4** [One-Shot, Buying] Given a probability distribution $f$ over future buyer pricelines $P^2$,

$$\mathsf{HOT}(f) = \max_{\beta^1 = \langle \vec{b}, 0 \rangle} \mathbb{E}_{P^2 \sim f} \big[ v(\mathrm{Buy}(\beta^1, P^2)) - \mathrm{Cost}(\mathrm{Buy}(\beta^1, P^2), P^2) \big] \quad (13)$$

**Hotel Bidding Problem, with Selling**  Although it is not possible for agents to sell TAC hotel auctions, one could imagine an analogous auction setup in which it were possible to sell goods as well as buy them.

**Definition 5.5** [One-Shot, Buying and Selling] Given a probability distribution $f$ over future buyer and seller pricelines $(P, \Pi)^2$,

$$\max_{\beta^1 = \langle \vec{b}, \vec{a} \rangle} \mathbb{E}_{(P,\Pi)^2 \sim f} \big[ v(\mathrm{Buy}(\beta^1, P^2) \ominus \mathrm{Sell}(\beta^1, \Pi^2)) - \mathrm{Cost}(\mathrm{Buy}(\beta^1, P^2), P^2) + \mathrm{Revenue}(\mathrm{Sell}(\beta^1, \Pi^2), \Pi^2) \big]$$
$$(14)$$

subject to $\mathrm{Buy}(\beta^1, P^2) \geq \mathrm{Sell}(\beta^1, \Pi^2)$, for all $(P, \Pi)^2$.

**Bidding Problem**  Finally, we present (a slight generalization of) the TAC bidding problem by combining the four previous stochastic optimization problems into one. This abstract problem models bidding to buy and sell goods both via continuously-clearing and one-shot second-price pseudo-auctions, as follows:

**Definition 5.6** [Bidding Under Uncertainty] Given a set of current buyer and seller pricelines $(P, \Pi)^1$ and a probability distribution $f$ over future buyer and seller pricelines $(P, \Pi)^2$,

$$\mathsf{BID}(f) =$$

$$\max_{Q^1, R^1 \in \mathbb{Z}^n, \beta^1 = \langle \vec{b}, \vec{a} \rangle} \mathbb{E}_{(P,\Pi)^2 \sim f} \Big[ \max_{Q^2, R^2 \in \mathbb{Z}^n} v((Q^1 \oplus Q^2) \ominus (R^1 \oplus R^2) \oplus \mathrm{Buy}(\beta^1, P^2) \ominus \mathrm{Sell}(\beta^1, P^2))$$
$$- \big( \mathrm{Cost}(Q^1, P^1) + \mathrm{Cost}(Q^1 \oplus Q^2, P^2) - \mathrm{Cost}(Q^1, P^2) + \mathrm{Cost}(\mathrm{Buy}(\beta^1, P^2), P^2) \big)$$
$$+ \big( \mathrm{Revenue}(R^1, \Pi^1) + \mathrm{Revenue}(R^1 \oplus R^2, \Pi^2) - \mathrm{Revenue}(R^1, \Pi^2) + \mathrm{Revenue}(\mathrm{Sell}(\beta^1, \Pi^2), \Pi^2) \big) \Big]$$
$$(15)$$

subject to $Q^1 \supseteq R^1$ and $Q^1 \oplus Q^2 \supseteq R^1 \oplus R^2$ and $\text{Buy}(\beta^1, P^2) \geq \text{Sell}(\beta^1, \Pi^2)$, for all $(P, \Pi)^2$.

Once again, this bidding problem is (i) stochastic: it takes as input a stochastic model of future prices; (ii) global: it seamlessly integrates flight, hotel, and entertainment bidding decisions; and (iii) dynamic: it facilitates simultaneous reasoning about current and future stages of the game.

Next, we describe various heuristic approaches to solving the problem of bidding under uncertainty.

## 5.3 Bidding Heuristics

In this section, we discuss two heuristic solutions to the bidding problem: specifically, the expected value method (EVM), an approach that collapses stochastic information, and sample average approximation (SAA), an approach that exploits stochastic information and characterizes RoxyBot-06.

### 5.3.1 Expected Value Method

The *expected value method* (Birge & Louveaux, 1997) is a standard way of approximating the solution to a stochastic optimization problem. First, the given distribution is collapsed into a point estimate (e.g., the mean); then, a solution to the corresponding deterministic optimization problem is output as an approximate solution to the original stochastic optimization problem. Applying this idea to the problem of bidding under uncertainty yields:

**Definition 5.7** [Expected Value Method] Given a probability distribution $f$ over buyer and seller pricelines, with expected values $\bar{P}^2$ and $\bar{\Pi}^2$, respectively,

$$
\begin{aligned}
\text{BID\_EVM}(\bar{P}^2, \bar{\Pi}^2) = & \\
\max_{Q^1, R^1 \in \mathbb{Z}^n, \beta^1 = \langle \vec{b}, \vec{a} \rangle, Q^2, R^2 \in \mathbb{Z}^n} & \ v((Q^1 \oplus Q^2) \ominus (R^1 \oplus R^2) \oplus (\text{Buy}(\beta^1, \bar{P}^2) \ominus \text{Sell}(\beta^1, \bar{P}^2)) \\
& - \big(\text{Cost}(Q^1, P^1) + \text{Cost}(Q^1 \oplus Q^2, \bar{P}^2) - \text{Cost}(Q^1, \bar{P}^2) + \text{Cost}(\text{Buy}(\beta^1, \bar{P}^2), \bar{P}^2)\big) \\
& + \big(\text{Revenue}(R^1, \Pi^1) + \text{Revenue}(R^1 \oplus R^2, \bar{\Pi}^2) - \text{Revenue}(R^1, \bar{\Pi}^2) + \text{Revenue}(\text{Sell}(\beta^1, \bar{\Pi}^2), \bar{\Pi}^2)\big)
\end{aligned}
\tag{16}
$$

subject to $Q^1 \supseteq R^1$ and $Q^1 \oplus Q^2 \supseteq R^1 \oplus R^2$.

In practice, without full knowledge of the distribution $f$, we cannot implement the expected value method; in particular, we cannot compute $\bar{P}^2$ or $\bar{\Pi}^2$ so we cannot solve $\text{BID\_EVM}(\bar{P}^2, \bar{\Pi}^2)$ exactly. We can, however, solve an approximation of this problem in which the expected buyer and seller pricelines $\bar{P}^2$ and $\bar{\Pi}^2$ are replaced by an average scenario $(\hat{P}^2, \hat{\Pi}^2)$ (i.e., average buyer and seller pricelines), defined as follows:

$$
\hat{P}^2 = \frac{1}{S} \sum_{i=1}^{S} P_i^2, \qquad \hat{\Pi}^2 = \frac{1}{S} \sum_{i=1}^{S} \Pi_i^2.
$$

---

**Algorithm 5** EVM$(G, N, f, S)$

1: sample $S$ scenarios $(P, \Pi)_1^2, \ldots, (P, \Pi)_S^2 \sim f$
2: $\beta \Leftarrow$ BID_EVM $\left( \sum_{i=1}^S P_i^2, \sum_{i=1}^S \Pi_i^2 \right)$
3: **return** $\beta$

---

### 5.3.2 SAMPLE AVERAGE APPROXIMATION

Like the expected value method, *sample average approximation* is an intuitive way of approximating the solution to a stochastic optimization problem. The idea is simple: (i) generate a set of sample scenarios, and (ii) solve an approximation of the problem that incorporates only the sample scenarios. Applying the SAA heuristic (see Algorithm 6) involves solving the following approximation of the bidding problem:

**Definition 5.8** [Sample Average Approximation] Given a set of $S$ scenarios, $(P, \Pi)_1^2, \ldots, (P, \Pi)_S^2 \sim f$,

$$
\text{BID\_SAA}((P, \Pi)_1^2, \ldots, (P, \Pi)_S^2) =
$$

$$
\max_{Q^1, R^1 \in \mathbb{Z}^n, \beta^1 = \langle \vec{b}, \vec{a} \rangle} \sum_{i=1}^S \max_{Q^2, R^2 \in \mathbb{Z}^n} v((Q^1 \oplus Q^2) \ominus (R^1 \oplus R^2) \oplus (\text{Buy}(\beta^1, P_i^2) \ominus \text{Sell}(\beta^1, P_i^2))
$$

$$
- \left( \text{Cost}(Q^1, P^1) + \text{Cost}(Q^1 \oplus Q^2, P_i^2) - \text{Cost}(Q^1, P_i^2) + \text{Cost}(\text{Buy}(\beta^1, P_i^2), P_i^2) \right)
$$

$$
+ \left( \text{Revenue}(R^1, \Pi^1) + \text{Revenue}(R^1 \oplus R^2, \Pi_i^2) - \text{Revenue}(R^1, \Pi_i^2) + \text{Revenue}(\text{Sell}(\beta^1, \Pi_i^2), \Pi_i^2) \right)
$$

$$(17)$$

subject to $Q^1 \supseteq R^1$ and $Q^1 \oplus Q^2 \supseteq R^1 \oplus R^2$.

---

**Algorithm 6** SAA$(G, N, f, S)$

1: sample $S$ scenarios $(P, \Pi)_1^2, \ldots, (P, \Pi)_S^2 \sim f$
2: $\beta \Leftarrow$ BID_SAA$((P, \Pi)_1^2, \ldots, (P, \Pi)_S^2)$
3: **return** $\beta$

---

Using the theory of large deviations, Ahmed and Shapiro (2002) establish the following result: as $S \to \infty$, the probability that an optimal solution to the sample average approximation of a stochastic program with integer recourse is an optimal solution to the original stochastic optimization problem approaches 1 exponentially fast. Given hard time and space constraints, however, it is not always possible to sample sufficiently many scenarios to infer any reasonable guarantees about the quality of a solution to a sample average approximation. Hence, we propose a modified SAA heuristic, in which SAA is fed some tailor-made "important" scenarios, and we apply this idea to the bidding problem.

### 5.3.3 MODIFIED SAMPLE AVERAGE APPROXIMATION

The bids that SAA places are sample prices that appear in its scenarios. SAA never bids higher on any good than its highest sampled price, because as far as it knows, bidding that price is enough to win that good in all scenarios. However, there is some chance that the highest sampled price falls below the clearing price. Let us compute this probability in the case of a single-unit auction, or a *uniform-price* multi-unit auction: i.e., one in which all units of the good being auctioned off clear at the same price.

Let $F$ denote the cumulative distribution function over the predicted prices, let $f$ denote the corresponding density function, and let $G$ denote the cumulative distribution function over the clearing prices. Using this notation, the term $1 - G(x)$ is the probability the clearing price is greater than $x$. Further, let $X$ be a random variable that represents the highest value among $S$ sample price predictions. Then $P(X \leq x) = F(x)^S$ is the probability that all $S$ samples (and hence the highest among them) are less than $x$; and $P(X = x) = (F(x)^S)' = S(F(x))^{S-1}f(x)$ is the probability that the highest value among the $S$ samples equals $x$. Putting these two terms together—namely, the probability the highest sample price prediction is exactly $x$, and the probability the clearing price is greater than $x$—we can express the probability the highest of SAA's sample price predictions is less than the clearing price as follows:

$$\int_{-\infty}^{\infty} S(F(x))^{S-1}f(x)(1 - G(x))dx \tag{18}$$

Assuming perfect prediction (so that $G = F$), this complex expression simplies as follows:

$$\int_{-\infty}^{\infty} S(F(x))^{S-1}f(x)(1 - F(x))dx$$
$$= S\int_{-\infty}^{\infty}(F(x))^{S-1}f(x)dx \; - \; S\int_{-\infty}^{\infty}(F(x))^S f(x)dx$$
$$= S\left[\frac{(F(x))^S}{S}\right]_{-\infty}^{\infty} - S\left[\frac{(F(x))^{S+1}}{S+1}\right]_{-\infty}^{\infty}$$
$$= \frac{1}{S+1}$$

Hence, the probability that all SAA's sample price predictions are less than the clearing price is $1/(S+1)$. In particular, assuming perfect prediction and that the clearing prices in the TAC hotel auctions are independent, the probability that an SAA agent with 49 scenarios bidding in TAC Travel has any chance of winning all eight hotels (i.e., the probability that a sample price in at least one of its scenarios is greater than the clearing price) is only $\left(1 - \frac{1}{49+1}\right)^8 = 0.98^8 \approx 0.85$.

To remedy this situation, we designed and implemented a simple variant of SAA in RoxyBot-06. The SAA* heuristic (see Algorithm 7) is a close cousin of SAA, the only difference arising in their respective scenario sets. Whereas SAA samples $S$ scenarios, SAA* samples only $S - |N|$ scenarios, where $|N| = \sum_g N_g$. SAA* creates an additional $|N|$ scenarios as follows: for each unit $k$ of each good $g \in G$, it sets the price of the $k$th unit of good $g$ to the upper limit of its range of possible prices and, after conditioning on this price setting, it sets the prices of the other goods to their mean values. Next, we describe experiments with a test suite of bidding heuristics, including SAA and SAA*, in a controlled testing environment.

---

**Algorithm 7** SAA*$(G, N, f, S)$

---

**Require:** $S \geq |N|$

1: hard-code $|N|$ scenarios $(P, \Pi)_1^2, \ldots, (P, \Pi)_{|N|}^2$
2: sample $S - |N|$ scenarios $(P, \Pi)_{|N|+1}^2, \ldots, (P, \Pi)_S^2 \sim f$
3: $\beta \Leftarrow$ BID_SAA$((P, \Pi)_1^2, \ldots, (P, \Pi)_S^2)$
4: **return** $\beta$

---

| Agent | Predictions | Bids | On |
|-------|-------------|------|-----|
| SMU | Average scenario | Marginal utilities | All goods |
| AMU | $S$ scenarios | Calculates marginal utilities in each scenario Bids average marginal utilities across scenarios | All goods |
| TMU | Average scenario | Marginal utilities | Goods in a target set |
| BE | $S$ scenarios | Best of $S$ TMU solutions | Goods in a target set |
| TMU* | Average scenario | Marginal utilities, assuming only goods in a target set are available | Goods in a target set |
| BE* | $S$ scenarios | Best of $S$ TMU*solutions | Goods in a target set |

Table 3: Marginal-utility-based agents. The marginal utility of a good is defined as the incremental utility that can be achieved by winning that good, relative to the utility of the set of goods already held.

## 5.4 Summary

In this section, we developed a series of bidding problems, and heuristics solutions to those problems, that captures the essence of bidding in the one-shot and continuously-clearing auctions that characterize TAC. The bulk of our presentation was deliberately abstract, so as to suggest that our problems and their solutions are applicable well beyond the realm of TAC: e.g., to bidding for interdependent goods in separate eBay auctions. Still, it remains to validate our approach in other application domains.

## 6. Experiments

We close this paper with two sets of experimental results, the first in a controlled testing environment, and the second the results from the final round of the 2006 TAC Travel competition. The combined strategy of hotel price prediction via SimAA and bid optimization via SAA emerged victorious in both settings.

### 6.1 Controlled Experiments

To some extent at least, our approach to bidding has been validated by the success of RoxyBot-06 in TAC-06. Nonetheless, we ran simulations in a controlled testing environment to further validate our approach. These results are reported by Lee (2007) and Greenwald et al. (2008), but we summarize them here as well.

We built a test suite of agents, all of which predict using RoxyBot-06's SimAA random mechanism with distribution. The agents differ in their bidding strategies; the possibilities include SAA,[8] SAA*, and the six marginal-utility-based heuristics studied by Wellman et al. (2007), and summarized in Table 3.

Our experiments were conducted in a TAC Travel-like setting, modified to remove any aspects of the game that would obscure a controlled study of bidding. Specifically, we eliminated flight and entertainment trading, and endowed all agents with eight flights in and eight flights out on each day. Further, we assumed all hotels closed after one round of bidding (i.e., hotel auctions are one-shot, so that the ensuing bid optimization problem adheres to Definition 5.4).

We designed two sets of experiments: one decision-theoretic and one game-theoretic. In the former, hotel clearing prices are the outcome of a simulation of simultaneous ascending auctions, but depend on the actual clients in each game, not a random sampling. (Our simulator is more informed than the individual agents.) In the latter, hotel clearing prices are determined by the bids the agents submit using the same mechanism as in TAC Travel: the clearing price is the 16th highest bid (or zero, if fewer than 16 bids are submitted).

We first ran experiments with 8 agents per game, but found that hotel prices were often zero: i.e., there was insufficient competition. We then changed the setup to include a random number of agents drawn from a binomial distribution with $n = 32$ and $p = 0.5$, with the requisite number of agents sampled uniformly with replacement from the set of possible agents. The agents first sample the number of competitors from the binomial distribution, and then generate scenarios assuming the sampled number of competitors.

Because of the game-theoretic nature of TAC, an individual agent's performance can depend heavily on the other agents included in the agent pool. In our experiments, we attempted to mitigate any artificial effects of the specific agents we chose to include in our pool by sampling agents from the pool to play each game, with replacement. Thus, an agent's average score from the games is a measure of the agent's performance against various combinations of opponents.

In Figures 3(a) and 3(b), we plot the mean scores obtained by each agent type in each setting, along with 95% confidence intervals. These averages were computed based on 1000 independent observations, obtained by playing 1000 games. Scores were averaged across agent types in each game to account for any game dependencies. SAAB and SAAT[9] are the best performing agents in the game-theoretic experiments and among the best in the decision-theoretic setting.

## 6.2 TAC 2006 Competition Results

Table 4 lists the agents entered in TAC-06 and Table 5 summarizes the outcome. The TAC-06 finals comprised 165 games over three days, with the 80 games on the last day weighted 1.5 times as much as the 85 over the first two days. On the first day of the finals, RoxyBot finished third, behind Mertacor and Walverine—the top scorers in 2005. As it happens, RoxyBot's optimization routine, which was designed for stochastic hotel and entertainment price predictions, was accidentally fed deterministic predictions (i.e., point price estimates) for entertainment. Moreover, these predictions were fixed, rather than adapted based on recent game history.

---

8. The particular implementation details explaining how RoxyBot-06 applied SAA in the TAC domain are relegated to Appendix A.

9. SAAB is SAA, and SAAT is a slight variant of SAA*. See the paper by Greenwald et al. (2008) for details.

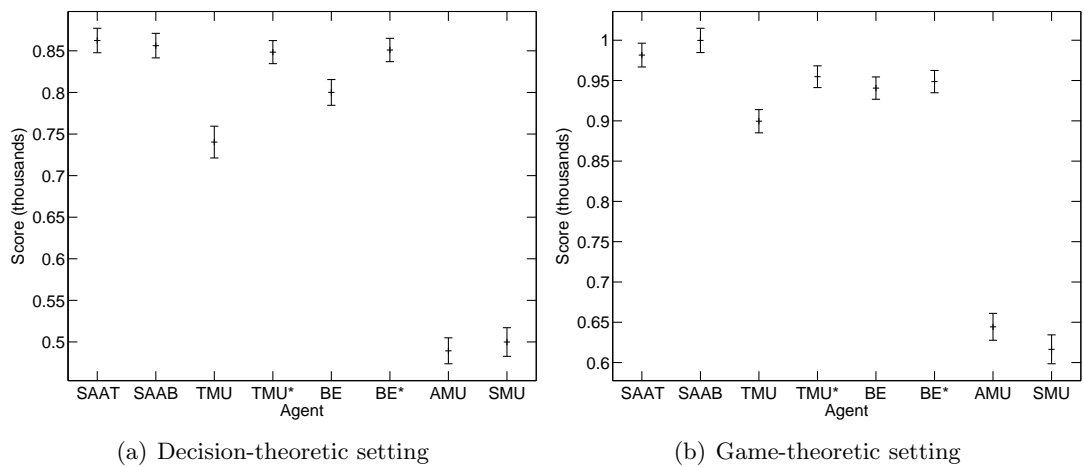

(a) Decision-theoretic setting       (b) Game-theoretic setting

Figure 3: Mean scores and confidence intervals.

On days 2 and 3, RoxyBot ran properly, basing its bidding in all auctions on stochastic information. Moreover, the agent was upgraded after day 1 to bid on flights not just once, but twice, during each minute. This enabled the agent to delay its bidding somewhat at the end of a game for flights whose prices are decreasing. No doubt this minor modification enabled RoxyBot to emerge victorious in 2006, edging out Walverine by a whisker, below the integer precision reported in Table 5. The actual margin was 0.22—a mere 22 parts in 400,000. Adjusting for control variates (Ross, 2002) spreads the top two finishers a bit further.[10]

| Agent | Affiliation | Reference |
|-------|-------------|-----------|
| 006 | Swedish Inst Comp Sci | Aurell et al., 2002 |
| kin_agent | U Macau | |
| L-Agent | Carnegie Mellon U | Sardinha et al., 2005 |
| Mertacor | Aristotle U Thessaloniki | Toulis et al., 2006; Kehagias et al., 2006 |
| RoxyBot | Brown U | Greenwald et al., 2003, 2004, 2005; Lee et al., 2007 |
| UTTA | U Tehran | |
| Walverine | U Michigan | Cheng et al., 2005; Wellman et al., 2005 |
| WhiteDolphin | U Southampton | He & Jennings, 2002; Vetsikas & Selman, 2002 |

Table 4: TAC-06 participants.

---

10. Kevin Lochner computed these adjustment factors using the method described by Wellman et al. (2007, ch. 8).

| Agent | Finals | Adjustment Factor |
|---|---|---|
| RoxyBot | 4032 | −5 |
| Walverine | 4032 | −17 |
| WhiteDolphin | 3936 | −2 |
| 006 | 3902 | −27 |
| Mertacor | 3880 | −16 |
| L-Agent | 3860 | 7 |
| kin_agent | 3725 | 0 |
| UTTA | 2680 | −14 |

Table 5: TAC-06 final scores, with adjustment factors based on control variates.

Mean scores, utilities, and costs (with 95% confidence intervals) for the last day of the TAC-06 finals (80 games) are plotted in Figure 4 and detailed statistics are tabulated in Table 6. There is no single metric such as low hotel or flight costs that is responsible for RoxyBot's success. Rather its success derives from the right balance of contradictory goals. In particular, RoxyBot incurs high hotel and mid-range flight costs while achieving mid-range trip penalty and high event profit.[11]

Let us compare RoxyBot with two closest rivals: Walverine and WhiteDolphin. Comparing to Walverine first, Walverine bids lower prices (by 55) on fewer hotels (49 less), yet wins more (0.8) and wastes less (0.42). It would appear that Walverine's hotel bidding strategy outperforms RoxyBot's, except that RoxyBot earns a higher hotel bonus (15 more). RoxyBot also gains an advantage by spending 40 less on flights and earning 24 more in total entertainment profit.

A very different competition takes place between RoxyBot and WhiteDolphin. WhiteDolphin bids lower prices (120 less) on more hotels (by 52) than RoxyBot. RoxyBot spends much more (220) on hotels than WhiteDolphin but makes up for it by earning a higher hotel bonus (by 96) and a lower trip penalty (by 153). It seems that WhiteDolphin's strategy is to minimize costs even if that means sacrificing utility.

### 6.3 Summary

As already noted, TAC Travel bidding, viewed as an optimization problem, is an $n$-stage decision problem. We solve this $n$-stage decision problem as a sequence of 2-stage decision problems. The controlled experiments reported in this section establish that our bidding strategy, SAA, is the best in our test suite in the setting for which it was designed, with only 2 stages. The TAC competition results establish that this strategy is also effective in an $n$-stage setting.

## 7. Collective Behavior

The hotel price prediction techniques described in Section 4.2 are designed to compute (or at least approximate) competitive equilibrium prices without full knowledge of the client pop-ulation. In this section, we assume this knowledge and view the output of the tâtonnement and SimAA calculations not as predictions but as ground truth. We compare the actual prices in the final games to this ground truth in respective years since 2002 to determine whether TAC market prices resemble CE prices. What we find is depicted in Figure 5. Because of the nature of our methods, these calculations pertain to hotel prices only.

---

11. An agent suffers trip penalties to the extent that it assigns its clients packages that differ from their preferred.

|                     | Rox   | Wal   | Whi   | SIC   | Mer   | L-A   | kin   | UTT   |
|---------------------|-------|-------|-------|-------|-------|-------|-------|-------|
| # of Hotel Bids     | 130   | 81    | 182   | 33    | 94    | 58    | 15    | 24    |
| Average of Hotel Bids | 170 | 115   | 50    | 513   | 147   | 88    | 356   | 498   |
| # of Hotels Won     | 15.99 | 16.79 | 23.21 | 13.68 | 18.44 | 14.89 | 15.05 | 9.39  |
| Hotel Costs         | 1102  | 1065  | 882   | 1031  | 902   | 987   | 1185  | 786   |
| # of Unused Hotels  | 2.24  | 1.82  | 9.48  | 0.49  | 4.86  | 1.89  | 0.00  | 0.48  |
| Hotel Bonus         | 613   | 598   | 517   | 617   | 590   | 592   | 601   | 424   |
| Trip Penalty        | 296   | 281   | 449   | 340   | 380   | 388   | 145   | 213   |
| Flight Costs        | 4615  | 4655  | 4592  | 4729  | 4834  | 4525  | 4867  | 3199  |
| Event Profits       | 110   | 26    | 6     | -6    | 123   | -93   | -162  | -4    |
| Event Bonus         | 1470  | 1530  | 1529  | 1498  | 1369  | 1399  | 1619  | 996   |
| Total Event Profits | 1580  | 1556  | 1535  | 1492  | 1492  | 1306  | 1457  | 992   |
| Average Utility     | 9787  | 9847  | 9597  | 9775  | 9579  | 9604  | 10075 | 6607  |
| Average Cost        | 5608  | 5693  | 5468  | 5765  | 5628  | 5605  | 6213  | 3989  |
| Average Score       | 4179  | 4154  | 4130  | 4010  | 3951  | 3999  | 3862  | 2618  |

Table 6: 2006 Finals, Last day. Tabulated Statistics. We omit the first two days because agents can vary across days, but cannot vary within. Presumably, the entries on the last day are the teams' preferred versions of the agents.

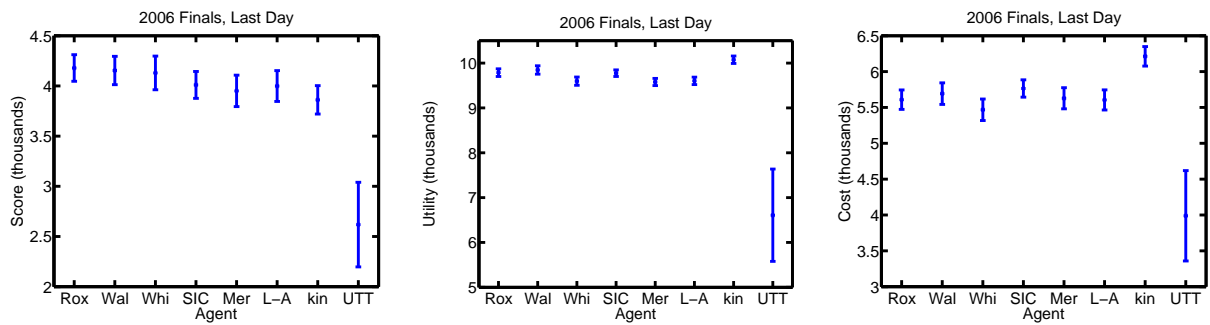

Figure 4: 2006 Finals, Last day. Mean scores, utilities, and costs, and 95% confidence intervals.

The results are highly correlated on both metrics (Euclidean distance and EVPP). We observe that the accuracy of CE price calculations has varied from year to year. 2003 was the year in which TAC Supply Chain Management (SCM) was introduced. Many participants diverted their attention away from Travel towards SCM that year, perhaps leading to degraded performance in Travel. Things seem to improve in 2004 and 2005. We cannot explain the setback in 2006, except by noting that performance is highly dependent on the particular agent pool, and in 2006 there were fewer agents in that pool.

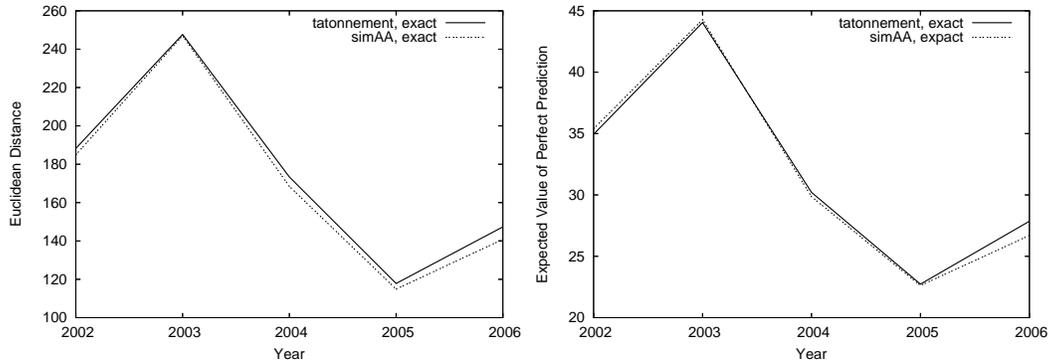

Figure 5: A comparison of the actual (hotel) prices to the output of competitive equilibrium price calculations in the final games since 2002. The label "exact" means: full knowledge of the client population.

## 8. Conclusion

The foremost aim of trading agent research is to develop a body of techniques for effective design and analysis of trading agents. Contributions to trading agent design include the invention of trading strategies, together with models and algorithms for realizing their computation and methods to measure and evaluate the performance of agents characterized by those strategies. Researchers seek both specific solutions to particular trading problems and general principles to guide the development of trading agents across market scenarios. This paper purports to contribute to this research agenda. We described the design and implementation of RoxyBot-06, an able trading agent as demonstrated by its performance in TAC-06.

Although automated trading in electronic markets has not yet fully taken hold, the trend is well underway. Through TAC, the trading agent community is demonstrating the potential for autonomous bidders to make pivotal trading decisions in a most effective way. Such agents offer the potential to accelerate the automation of trading more broadly, and thus shape the future of commerce.

## Acknowledgments

This paper extends the work of Lee et al. (2007). The material in Section 5.1 is based on the book by Wellman et al. (2007). We are grateful to several anonymous reviewers whose constructive criticisms enhanced the quality of this work. This research was supported by NSF Career Grant #IIS-0133689.

## Appendix A. TAC Bidding Problem: SAA

The problem of bidding in the simultaneous auctions that characterize TAC can be formulated as a two-stage stochastic program. In this appendix, we present the implementation details of the integer linear program (ILP) encoded in RoxyBot-06 that approximates an optimal solution to this stochastic program.[12]

We formulate this ILP assuming current prices are known, and future prices are uncertain in the first stage but revealed in the second stage. Note that whenever prices are known, it suffices for an agent to make decisions about the quantity of each good to buy, rather than about bid amounts, since choosing to bid an amount that is greater than or equal to the price of a good is equivalent to a decision to buy that good.

Unlike in the main body of the paper, this ILP formulation of bidding in TAC assumes linear prices. Table 7 lists the price constants and decision variables for each auction type. For hotels, the only decisions pertain to buy offers; for flights, the agent decides how many tickets to buy now and how many to buy later; for entertainment events, the agent chooses sell quantities as well as buy quantities.

| Hotels | Price | Variable (bid) |
|--------|-------|----------------|
| bid now | $\mathcal{Y}_{as}$ | $\phi_{apq}$ |

| Flights and Events | Price | Variable (qty) |
|--------------------|-------|----------------|
| buy now | $\mathcal{M}_a$ | $\mu_a$ |
| buy later | $\mathcal{Y}_{as}$ | $v_{as}$ |

| Events | Price | Variable (qty) |
|--------|-------|----------------|
| sell now | $\mathcal{N}_a$ | $\nu_a$ |
| sell later | $\mathcal{Z}_{as}$ | $\zeta_{as}$ |

Table 7: Auction types and associated price constants and decision variables.

### A.1 Index Sets

$a \in A$ indexes the set of goods, or auctions.

$a_f \in A_f$ indexes the set of flight auctions.

$a_h \in A_h$ indexes the set of hotel auctions.

$a_e \in A_e$ indexes the set of event auctions.

$c \in C$ indexes the set of clients.

$p \in P$ indexes the set of prices.

---

12. The precise formulation of RoxyBot-06's bidding ILP appears in the paper by Lee et al. (2007). The formulation here is slightly simplified, but we expect it would perform comparably in TAC. The key differences are in flight and entertainment bidding.

$q \in Q$ indexes the set of quantities
(i.e., the units of each good in each auction).

$s \in S$ indexes the set of scenarios.

$t \in T$ indexes the set of trips.

## A.2 Constants

$\mathcal{G}_{at}$ indicates the quantity of good $a$ required to complete trip $t$.

$\mathcal{M}_a$ indicates the current buy price of $a_f, a_e$.

$\mathcal{N}_a$ indicates the current sell price of $a_e$.

$\mathcal{Y}_{as}$ indicates the future buy price of $a_f, a_h, a_e$ in scenario $s$.

$\mathcal{Z}_{as}$ indicates the future sell price of $a_e$ in scenario $s$.

$\mathcal{H}_a$ indicates the hypothetical quantity won of hotel $a_h$.

$\mathcal{O}_a$ indicates the quantity of good $a$ the agent owns.

$\mathcal{U}_{ct}$ indicates client $c$'s value for trip $t$.

## A.3 Decision Variables

$\Gamma = \{\gamma_{cst}\}$ is a set of boolean variables indicating whether or not client $c$ is allocated trip $t$ in scenario $s$.

$\Phi = \{\phi_{apq}\}$ is a set of boolean variables indicating whether to bid price $p$ on the $q$th unit of $a_h$.

$M = \{\mu_a\}$ is a set of integer variables indicating how many units of $a_f, a_e$ to buy now.

$N = \{\nu_a\}$ is a set of integer variables indicating how many units of $a_e$ to sell now.

$Y = \{v_{as}\}$ is a set of integer variables indicating how many units of $a_f, a_e$ to buy later in scenario $s$.

$Z = \{\zeta_{as}\}$ is a set of integer variables indicating how many units of $a_e$ to sell later in scenario $s$.

## A.4 Objective Function

$$\max_{\Gamma, \Phi, M, N, Y, Z} \sum_S \left( \overbrace{\sum_{C,T} \mathcal{U}_{ct} \gamma_{cts}}^{trip\ value} - \sum_{A_f} \overbrace{\left( \overbrace{\mathcal{M}_a \mu_a}^{current} + \overbrace{\mathcal{Y}_{as} v_{as}}^{future} \right)}^{flight\ cost} - \overbrace{\sum_{A_h, Q, p \geq \mathcal{Y}_{as}} \mathcal{Y}_{as} \phi_{apq}}^{hotel\ cost} + \quad (19) \right.$$

$$\sum_{A_e} \left( \overbrace{\underbrace{\mathcal{N}_a \nu_a}_{current} + \underbrace{\mathcal{Z}_{as}\zeta_{as}}_{future}}^{event\ revenue} - \overbrace{\underbrace{\mathcal{M}_a \mu_a}_{current} - \underbrace{\mathcal{Y}_{as} \upsilon_{as}}_{future}}^{event\ cost} \right) \right)$$

## A.5 Constraints

$$\sum_T \gamma_{cst} \leq 1 \quad \forall c \in C, s \in S \tag{20}$$

$$\overbrace{\sum_{C,T} \gamma_{cst} \mathcal{G}_{at}}^{allocation} \leq \overbrace{\mathcal{O}_a}^{own} + \overbrace{(\mu_a + \upsilon_{as})}^{buy} \quad \forall a \in A_f, s \in S \tag{21}$$

$$\overbrace{\sum_{C,T} \gamma_{cst} \mathcal{G}_{at}}^{allocation} \leq \overbrace{\mathcal{O}_a}^{own} + \overbrace{\sum_{Q, p \geq \mathcal{Y}_{as}} \phi_{apq}}^{buy} \quad \forall a \in A_h, s \in S \tag{22}$$

$$\overbrace{\sum_{C,T} \gamma_{cst} \mathcal{G}_{at}}^{allocation} \leq \overbrace{\mathcal{O}_a}^{own} + \left( \overbrace{\mu_a + \upsilon_{as}}^{buy} \right) - \left( \overbrace{\nu_a + \zeta_{as}}^{sell} \right)$$
$$\forall a \in A_e, s \in S \tag{23}$$

$$\sum_{P,Q} \phi_{apq} \geq \mathcal{H}_a \quad \forall a \in A_h \tag{24}$$

$$\sum_P \phi_{apq} \leq 1 \quad \forall a \in A_h, q \in Q \tag{25}$$

Equation (20) limits each client to one trip in each scenario. Equation (21) prevents the agent from allocating flights that it does not own or buy. Equation (22) prevents the agent from allocating hotels that it does not own or buy. Equation (23) prevents the agent from allocating event tickets that it does not own or buy and not sell. Equation (24) ensures the agent bids on at least HQW units in each hotel auction. Equation (25) prevents the agent from placing more than one buy offer per unit in each hotel auction.

An agent might also be constrained not to place sell offers on more units of each good than it owns, and/or not to place buy (sell) offers for more units of each good than the market supplies (demands).

Note that there is no need to explicitly enforce the bid monotonicity constraints in this ILP formulation:

- "Buy offers must be nonincreasing in k, and sell offers nondecreasing."

  The ILP does not need this constraint because prices are assumed to be linear. In effect, the only decisions the ILP makes are how many units of each good to bid on. Hence, the bids (10, 15, 20) and (20, 15, 10) are equivalent.

- "An agent may not offer to sell for less than the price it is willing to buy."

The ILP would not choose to place both a buy offer and a sell offer on a good if the buy price of that good exceeds the sell price, because that would be unprofitable.